\documentclass[jap,twocolumn,amsmath,amssymb]{revtex4-1}

\usepackage{graphicx}
\usepackage{dcolumn}
\usepackage{bm}

\begin{document}

\title{Fluctuation-induced tunneling conduction through RuO$_2$ nanowire contacts}

\author{Yong-Han Lin}
\email{yonghanlin@gmail.com}
\affiliation{Institute of Physics, National Chiao Tung University, Hsinchu 30010, Taiwan}

\author{Juhn-Jong Lin}
\email{jjlin@mail.nctu.edu.tw}
\affiliation{Institute of Physics, National Chiao Tung University, Hsinchu 30010, Taiwan}
\affiliation{Department of Electrophysics, National Chiao Tung University, Hsinchu 30010, Taiwan}

\date{\today}

\begin{abstract}

A good understanding of the electronic conduction processes through nanocontacts is a crucial step for the implementation of functional nanoelectronic devices. We have studied the current-voltage ($I$-$V$) characteristics of nanocontacts between single metallic RuO$_2$ nanowires (NWs) and contacting Au electrodes which were pre-patterned by simple photolithography. Both the temperature behavior of contact resistance in the
low-bias voltage ohmic regime and the $I$-$V$ curves in the high-bias voltage non-ohmic regime have been investigated. We found that the electronic conduction processes in the wide temperature interval 1--300 K can be well described by the fluctuation-induced tunneling (FIT) conduction theory. Taken together with our previous work (Lin {\it et al.}, Nanotechnology {\bf 19}, 365201 (2008)) where the nanocontacts were fabricated by delicate electron-beam lithography, our study demonstrates the general validity of the FIT model in characterizing electronic nanocontacts.

\end{abstract}



\maketitle

\section{Introduction}

Nanometer-scale devices based on different kinds of (metallic, semiconducting, magnetic, etc.) nanowires (NWs) have widely been demonstrated in various potential applications, such as nanoelectronics, nano-electromechanical systems, optoelectronics, and spintronics.\cite{1D} To characterize intrinsic NW properties as well as to interrogate the performance of a nanodevice, how to successfully attach submicron electrical contacts to individual NWs is a crucial step. Until recently, the charge conduction properties of electronic nanocontacts and their possible effects on device operations have often been overlooked, in particular in the case of metallic NWs.\cite{WangAPL04,MarziJAP04,LinNT08} Obviously, poor knowledge about such properties could lead to a less quantitative description or even a misinterpretation of the experimental data.\cite{ChiuNT09} Besides, ignorance of possible degradative effects due to the nanocontacts might cause damages to nanoelectronic circuitries. In this regard, a good understanding of the electronic conduction processes through a nanocontact is of both scientific and technological importance.

In a recent paper (Ref.~\onlinecite{LinNT08}), we have demonstrated that the temperature behavior of electronic contact resistances, $R_{\text{c}}(T)$, formed at the interfaces between various metallic (RuO$_2$, IrO$_2$, and Sn-doped In$_2$O$_{3-x}$) NWs and lithography-patterned submicron Cr/Au electrodes can be well described by the fluctuation-induced tunneling (FIT) conduction model proposed by Sheng and co-workers.\cite{ShengPRL78,ShengPRB80} In Ref.~\onlinecite{LinNT08}, the NWs were first disposed onto a SiO$_2$-capped Si wafer and the delicate electron-beam lithography and lift-off technique was applied to make the contacting electrodes in a two-probe configuration. Under proper conditions, one can intentionally make the nanocontact resistance $R_{\text{c}}$ much larger than both the metallic NW resistance and the contacting electrode resistance. Therefore, the measured device resistance $R(T)$\,$\simeq$\,$2R_{\text{c}}(T)$, where we have assumed that $R_{\text{c}}$ is the same for the two nanocontacts in a two-probe configuration because the two nanocontacts had been fabricated simultaneously and under the same conditions. The $R_{\text{c}}(T)$ curves between 1.5 and 300 K had been measured in the limit of zero bias voltage, where the current-voltage ($I$-$V$) characteristics were linear, i.e. ohmic. We explicitly demonstrated that $R_{\text{c}}$ increased with decreasing temperature, but with a rate much slower than that would be due to the thermal-activation-type conduction ($\propto$\,exp($T^\ast/T$), where $T^\ast$ is a characteristic temperature), especially at liquid-helium temperatures.

In this work, we aim to further test the validity of the FIT conduction model by employing an alternative nanodevice contact configuration. Here we fabricate individual NW devices with the NWs being directly placed on the surfaces of two pre-patterned Cr/Au electrodes which were made by the simple photolithography. In particular, the $I$-$V$ curves in both the low-bias voltage regime and the high-bias voltage (i.e. non-ohmic) regime have been measured. We found that our overall experimental results can still be well described by the FIT conduction theory, indicating the general importance of the thermal fluctuation effect in electronic nanocontacts.\cite{ShengPRL78,ShengPRB80} Our observations are reported below.

\section{Experimental method}

\begin{figure}[t]
\includegraphics[scale=0.1]{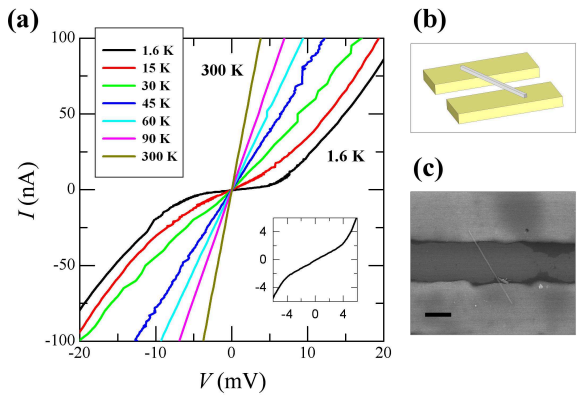}
\caption{ \label{fig_1} (a) Current-voltage curves at 1.6, 15, 30, 45, 60, 90, and 300 K for device A. The inset shows an expanded plot for the $I$-$V$ curve at 1.6 K. (b) A schematic of a contacting Au electrode pair bridged with a single NW. The NW is directly placed on the surfaces of the Au electrodes. (c) An SEM image of device A. The scale bar is 2 $\mu$m.}
\end{figure}

The RuO$_{2}$ NWs used in this study were synthesized by the thermal evaporation method, as described in Ref.~\onlinecite{LinNT08}. The as-synthesized NWs were dispersed by ultrasonic agitation into ethanol solution. Several droplets of the solution were dropped onto a Si substrate capped with a $\approx$\,500 nm SiO$_2$ layer. On the SiO$_2$ layer, an array of micron-sized electrode pairs consisting of two Cr/Au ($\approx$\,20/100 nm) fingers separated by $\sim$\,5 $\mu$m were pre-patterned by the standard photolithography and lift-off technique. The substrates were inspected by a scanning electron microscope (SEM), and those electrode pairs which were bridged with only one RuO$_2$ NW were selected for $I$-$V$ curve measurements. Electrical leads from the Cr/Au electrode pairs to macro Cu wires were connected with Ag paste. A Keithley model K-220 current source and a model K-617 electrometer were used. For measurements from 300 K down to 1.5 K, a standard $^4$He cryostat equipped with a calibrated silicon diode thermometer was utilized. Figures~\ref{fig_1}(b) and \ref{fig_1}(c) show a schematic diagram and an SEM image of an individual RuO$_2$ NW device, respectively. Two devices are studied in this work. Their sample parameters are listed in Table~\ref{table_1}.

\begin{table*}[t]
\caption{ \label{table_1} Values of relevant parameters for two RuO$_2$ NW devices. Device A (B) has a NW diameter of $\sim$\,70 ($\sim$\,100) nm. The junction (nanocontact) area $A$ was estimated from the SEM image and given by the product of the diameter of the NW and the average extent of the two Au contacting electrodes lying beneath the NW. The specific contact resistivity is defined by $\rho_{\text{c}}$\,=\,$R_{\text{c}}A$ and $\rho_{\text{c0}}=\rho_{\text{c}}$($T\rightarrow 0$ K). The intrinsic NW resistance for device A (B) is $\sim$\,2 ($\sim$\,1) k$\Omega$ and weakly dependent on temperature, as determined by the four-probe method in Ref.~\onlinecite{LinNT08}. The total Au electrode resistances are $\lesssim$\,10 $\Omega$.}
\begin{ruledtabular}
\begin{tabular}{ccccccccccc}
Device%
&Method%
&$R$(300 K)%
&$R_{\infty}$%
&$T_{1}$%
&$T_{0}$%
&$A$%
&$w$%
&$\phi_{0}$%
&$T_{1}$/$T_{0}$%
&$\rho_{\text{c0}}$%
\\
&%
&(k$\Omega$)%
&(k$\Omega$)%
&(K)%
&(K)%
&($\mu$m$^2$)%
&(nm)%
&(meV)%
&%
&(k$\Omega$~$\mu$m$^2$)%
\\
\hline
A &$R(T)$                       &37  &17.3 &159  &32.2 &0.166 &15.0 &1.7 &4.94 &401\\
  &$I_{\text{h}}(V_{\text{h}})$ &    &16.5 &206  &43.9 &      &13.7 &1.8 &4.69 &299\\
\hline
B &$R(T)$                       &118 &9.9  &1313 &234  &0.138 &10.5 &4.4 &5.61 &374\\
  &$I_{\text{h}}(V_{\text{h}})$ &    &20.4 &927  &198  &      &9.7  &3.6 &4.68 &304\\
\end{tabular}
\end{ruledtabular}
\end{table*}

We mention in passing that, to acquire a proper $R_{\text{c}}$ which is much larger than both the metallic NW resistance and the contacting electrode resistance, we simply cast the NW on top of the pre-fabricated contacting electrodes. By this method, high-resistance contacts are easily obtained.\cite{Muster00,ToimilMolares03} For semiconducting NWs,\cite{Gu01,Nagashima10} the tunnel barrier is an insulating layer inherently formed on the surface of the as-synthesized NW. However, for the metallic RuO$_2$ NWs,\cite{footnote1} the tunnel barrier separating the metallic NW and the contacting electrode is externally introduced,\cite{footnote2} and its exact origin needs further investigations.

\section{Results and discussion}

Figure~\ref{fig_1}(a) shows the measured $I$-$V$ characteristics for the device A at several temperatures between 1.6 and 300 K, as indicated in the caption to Fig.~\ref{fig_1}. It is seen that the $I$-$V$ curves are
symmetric about zero bias voltage at all measurement temperatures. At low bias voltages, the $I$-$V$ curves are linear, i.e. ohmic. However, when the bias voltage increases up to a certain level, a nonlinear behavior appears. As expected, the nonlinearity becomes progressively pronounced at lower temperatures. The inset of Fig.~\ref{fig_1}(a) indicates that the ohmic regime persists up to about 4 mV at 1.6 K. It should be noted that, even at our highest bias voltages and at the lowest measurement temperatures, the power dissipation in the NW device is still small ($\lesssim$\,2$\times$$10^{-9}$ W), and thus the electron overheating effect can be safely ignored. (The heat mainly dissipated at the two nanocontacts.)

\begin{figure}[t]
\includegraphics[scale=0.08]{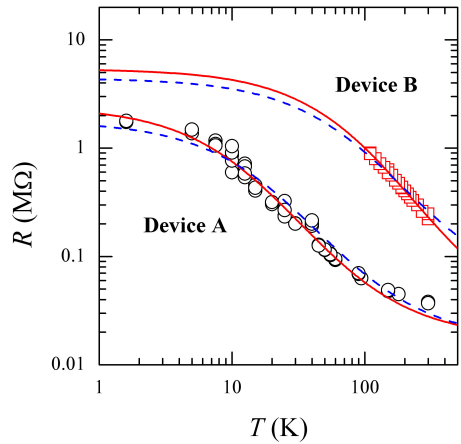}
\caption{ \label{fig_2} Zero-bias resistance as a function of temperature for two NW devices, as indicated. The symbols are the experimental data. The solid curves are the least-squares fits to Eq.~(\ref{RT}). The dashed curves are the theoretical predictions of the same equation but are plotted by directly substituting the $T_1$ and $T_0$ values extracted from $a(T)$ (Eq.~(\ref{a})). For clarity, the data for the device B have
been shifted up by multiplying a factor of 2.}
\end{figure}

From the $I$-$V$ curves measured at different temperatures, the temperature behavior of device resistance $R$ can be obtained. The $R$ is defined as the reciprocal of the slope of the $I$-$V$ curve around zero bias voltage. Figure~\ref{fig_2} shows a double logarithmic plot of the measured $R$ as a function of temperature for the devices A and B, as indicated. Similar to our previous finding in Ref.~\onlinecite{LinNT08}, one sees that the $R$ increases with decreasing temperature and eventually saturates at a few degrees of kelvin.\cite{footnote3} As has been demonstrated in Ref.~\onlinecite{LinNT08}, the sum of the individual RuO$_2$ NW resistance and the two Cr/Au electrode resistances is $\lesssim$\,1.5 k$\Omega$, which is $\sim$\,3 orders of magnitude smaller than $R$. Therefore, we may write $R(T)$\,$\simeq$\,$2R_{\text{c}}(T)$ for all temperatures, as mentioned.

The FIT model addresses the electronic conduction properties of a {\em mesoscopic} tunnel junction.\cite{ShengPRL78,ShengPRB80} More precisely, the FIT theory considers how the electron tunneling process is affected by temperature when the size of a tunnel junction formed between two {\em large} metallic regions becomes sufficiently {\em small}. The tunnel junction can be conceptually modeled as a potential barrier with a barrier height $\phi_0$, a barrier width $w$, and a lateral area $A$ which is the geometrical area at the closest contact between the two large metallic regions. In addition to the externally applied bias voltage, $V_{\text{a}}$, the FIT theory takes into account the effect of the thermal voltage noise, $V_{\text{th}}$, on the tunneling current across the junction. Sheng and coworkers realized that the $V_{\text{th}}$\,$\approx$\,$\sqrt{k_{\text{B}}T/C}$ could be large due to a small $A$, where $k_{\text{B}}$ is the Boltzmann constant, and $C$\,$\propto$\,$A$ is the capacitance of the junction. That is, a small junction area $A$ leads to a notable $V_{\text{th}}$. This large $V_{\text{th}}$ is of random sign with regard to $V_{\text{a}}$ and is shown to significantly modify the effective tunnel barrier seen by the electrons. As a result, the charge transport through the junction is profoundly affected.\cite{footnote4} Historically, the FIT theory has been formulated to explain the temperature dependence of resistance in a wide variety of metal-insulator composites and polymers, where a huge number of tunnel junctions is involved and an effective junction representing the whole network had to be assumed in the model.\cite{Wang94,Fisher04,Paschen95}

According to the FIT theory,\cite{ShengPRL78,ShengPRB80} at small applied electric fields where the $I$-$V$ characteristic is linear, the temperature dependent zero-bias resistance across a single small junction is given by
\begin{equation}
R(T)=R_{\infty}\,{\text{exp}}\left(\frac{T_{1}}{T_{0}+T}\right)~,%
\label{RT}%
\end{equation}
where $R_{\infty}$ is a parameter which depends only weakly on temperature, and $T_{1}$ and $T_{0}$ are characteristic temperatures defined as
\begin{equation}
T_1=\frac{8\varepsilon_0}{e^2k_{\text{B}}}\left(\frac{A \phi_0^2}{w}\right)~,%
\label{T1}%
\end{equation}
and
\begin{equation}
T_{0}=\frac{16\varepsilon_{0}\hbar}{\pi(2m)^{1/2}e^{2}k_{\text{B}}}
\left(\frac{A \phi_0^{3/2}}{w^2}\right)~,%
\label{T0}%
\end{equation}
where $\varepsilon_0$ is the permittivity of vacuum, $2\pi\hbar$ is the Planck's constant, and $m$ is the electronic mass. In Eq.~(\ref{RT}), $k_{\text{B}}T_1$ can be regarded as a measure of the energy required for an electron to cross the barrier, and $T_0$ is the temperature well below which thermal fluctuation effects become insignificant. Notice that, in the present work, the junction (nanocontact) area $A$ is experimentally determined by SEM and is not an adjustable parameter. For comparison, in previous studies of metal-insulator composites, polymers, and nanofibers,\cite{Wang94,Fisher04,Paschen95,Mandal11,AhnJMC10} the number of tunnel junctions and their individual areas are unknown and can not be experimentally determined.

For large applied electric fields where the $I$-$V$ curve becomes nonlinear, the FIT theory\cite{ShengPRL78,ShengPRB80} also gives an expression for the $I$-$V$ characteristic of the junction at the temperature $T$:
\begin{equation}
I_{\text{h}}(V_{\text{h}},T)=I_{\text{h,s}}\,%
{\text{exp}}\left[-a(T) \left(1-\frac{V_{\text{h}}}{V_{\text{h,c}}}\right)^{2}\,\right]~,%
~~~|V_{\text{h}}|<V_{\text{h,c}}~,%
\label{IV}%
\end{equation}
where the subscript `h' denotes the property at `high' bias voltages. The saturation current, $I_{\text{h,s}}$, and the critical voltage, $V_{\text{h,c}}$, are parameters which depend only weakly on temperature. $a(T)$ is a parameter which contains the dominant temperature effect on the $I$-$V$ curves and has the form
\begin{equation}
a(T)=\frac{T_{1}}{T_{0}+T}~.%
\label{a}%
\end{equation}
It should be noted that the ohmic regime corresponding to small applied voltages is not included in the derivation of Eq.~(\ref{IV}). Hence, the temperature dependent zero-bias resistance (Eq.~(\ref{RT})) cannot be
directly derived from Eq.~(\ref{IV}) by taking the $V_{\text{h}}$\,$\rightarrow$\,0 limit of $(dI_{\text{h}}/dV_{\text{h}})^{-1}$.

\begin{figure}[t]
\includegraphics[scale=0.1]{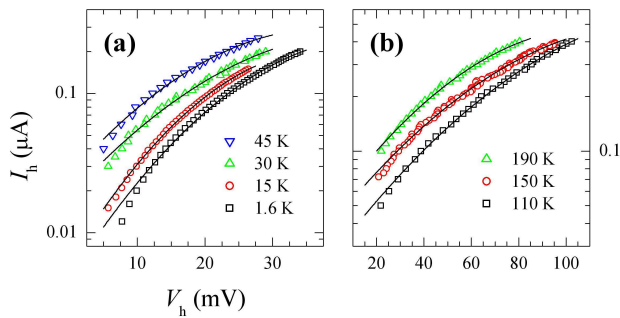}
\caption{ \label{fig_3} Current-voltage characteristics at high bias voltages for (a) device A and (b) device B at several temperatures, as indicated. The symbols are the experimental data. The solid curves are the least-squares fits to Eq.~(\ref{IV}).}
\end{figure}

Apparently, by measuring the $I$-$V$ curves at various temperatures, one has two different routes to extract the values of $T_{1}$ and $T_{0}$. The first route involves the low-bias voltage ohmic $I$-$V$ regime. The two values can be obtained by fitting the temperature behavior of $R$ to Eq.~(\ref{RT}). On the other hand, the second route involves the high-bias voltage non-ohmic $I$-$V$ regime. The value of $a(T)$ can first be extracted at each temperature by fitting the measured $I_{\text{h}}(V_{\text{h}})$ curve to Eq.~(\ref{IV}) (with the ohmic regime being excluded). Then, the values of $T_{1}$ and $T_{0}$ can be determined by fitting $a(T)$ to Eq.~(\ref{a}). These two complementary routes can provide a self-consistency check of the phenomenological FIT theory.

Return to Fig.~\ref{fig_2}. The symbols are the experimental data and the solid curves are the least-squares fits to Eq.~(\ref{RT}) with $R_\infty$, $T_1$ and $T_0$ as adjustable parameters. It is seen that the predictions of Eq.~(\ref{RT}) can well describe the experimental data. Our fitted $R_\infty$, $T_1$, and $T_0$ values are listed in Table~\ref{table_1}. (The dashed curves also describe the experimental data reasonably well, see further discussion below.) We stress again that the $R(T)$ behavior shown in Fig.~\ref{fig_2} is distinct from the thermal-activation-type conduction; it does not obey the $R$\,$\propto$\,exp($T^\ast/T$) law especially at liquid-helium temperatures (see, for example, the Fig.~1 of Ref.~\onlinecite{LinNT08}). The $R_\infty$, $T_1$, and $T_0$ values are in accord with those corresponding values obtained in Ref.~\onlinecite{LinNT08}.

\begin{figure}[t]
\includegraphics[scale=0.08]{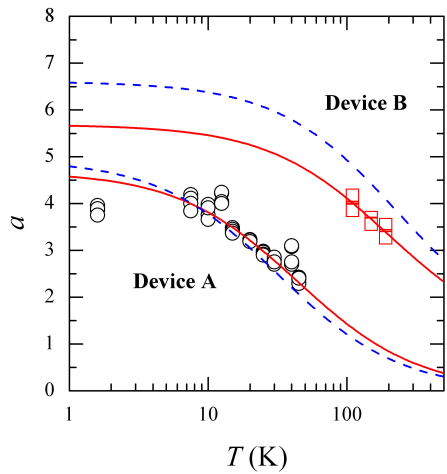}
\caption{ \label{fig_4} The parameter $a$ in Eq.~(\ref{IV}) as a function of temperature for devices A and B, as indicated. The solid curves are the least-squares fits to Eq.~(\ref{a}). The dashed curves are the theoretical predictions of the same equation but plotted by directly substituting the $T_1$ and $T_0$ values extracted from the $R(T)$ fits to Eq.~(\ref{RT}).}
\end{figure}

\begin{figure}[t]
\includegraphics[scale=0.08]{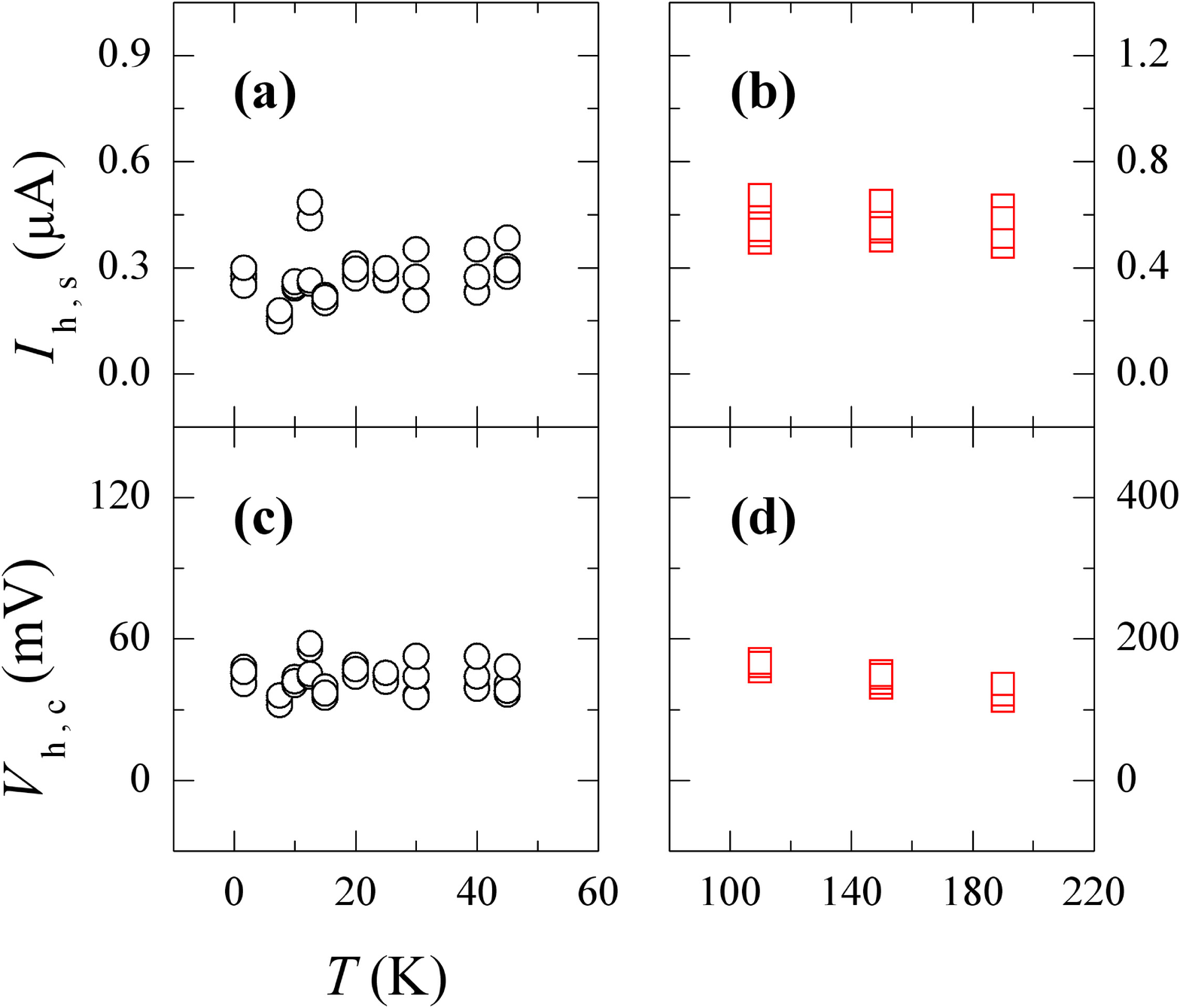}
\caption{ \label{fig_5} Variation of saturation current $I_{\text{h,s}}$ with temperature for (a) devices A and (b) device B, and variation of critical voltage $V_{\text{h,c}}$ with temperature for (c) devices A and (d) device B. Notice that these two parameters are temperature independent within our experimental uncertainties.}
\end{figure}

Figures~\ref{fig_3}(a) and \ref{fig_3}(b) show representative $I$-$V$ curves for the two devices in the nonlinear regime and at several temperatures, as indicated. The symbols are the experimental data and the solid curves are the least-squares fits to Eq.~(\ref{IV}) with $I_{\text{h,s}}$, $V_{\text{h,c}}$, and $a(T)$ as adjustable parameters. Clearly, the predictions of Eq.~(\ref{IV}) can well describe the experimental data in this non-ohmic regime. Our fitted values of $a$ as a function of temperature is plotted in Fig.~\ref{fig_4}, and our fitted values of $I_{\text{h,s}}$ and $V_{\text{h,c}}$ as a function of temperature are plotted in Fig.~\ref{fig_5}. Furthermore, the extracted values of $a(T)$ have been least-squares fitted to Eq.~(\ref{a}), as indicated by the solid curves in Fig.~\ref{fig_4}. The $T_1$ and $T_0$ values thus inferred are listed in Table~\ref{table_1}.

We now examine the self-consistency of the FIT theory by comparing the $T_{1}$ and $T_{0}$ values extracted from the two different manners, i.e. $R(T)$ and $I_{\text{h}}(V_{\text{h}})$. To do this, we plot in Fig.~\ref{fig_2} the predictions of Eq.~(\ref{RT}) by directly substituting in the $T_1$ and $T_0$ values inferred from $a(T)$ while leaving $R_{\infty}$ as the sole adjustable parameter. The results are indicated by the dashed curves in
Fig.~\ref{fig_2}. Clearly, the experimental data are fairly well reproduced in this manner. The $R_\infty$ value thus extracted is listed in Table~\ref{table_1}. On the other hand, we now substitute those $T_1$ and $T_0$ values extracted from $R(T)$ (Eq.~(\ref{RT})) into Eq.~(\ref{a}) to calculate $a(T)$. The calculated results are plotted as the dashed curves in Fig.~\ref{fig_4}, which is seen to well describe the $a(T)$ value for device A. (For device B, the discrepancy is somewhat larger, due to the lack of experimental data points at temperatures below 100 K.) In short, our $T_1$ and $T_0$ values inferred from the two methods differ by an amount of $\approx$\,30\%. This amount of difference is satisfactorily acceptable, and is somewhat smaller than that ($\approx$\,40\%) reported in a previous study of carbon nanotube network.\cite{KimSM01}

In the FIT theory, the barrier shape in the presence of an electric voltage across the junction is modeled by the image-force corrected rectangular barrier potential originally calculated by Simmons.\cite{SimmonsJAP63} It was found that different amounts of image-force correction result in different shapes of the barrier. With an increase in the electric voltage $V$ ($V$\,=\,$V_{\text{a}}$\,+\,$V_{\text{th}}$ is the total voltage), the barrier is effectively lowered and narrowed. Hence, there exists a (temperature independent) critical voltage, $V_{\text{h,c}}$, above which the peak value of the barrier is fully suppressed. Since the barrier is no longer visible to the electrons when $V$\,$>$\,$V_{\text{h,c}}$, a further increase in $V$ would only increase the energy of the electron but leave the barrier transmission factor unchanged (the transmission factor now becomes unity). This leads to a saturated tunneling current, $I_{\text{h,s}}$, which is insensitive to the voltage $V$ for $V$\,$>$\,$V_{\text{h,c}}$. Besides, it has been shown by numerical evaluation\cite{ShengPRB80} that $I_{\text{h,s}}$ depends only weakly on temperature. Therefore, the $I$-$V$ curves at all measurement temperatures are expected to saturate to a constant $I_{\text{h,s}}$. Indeed, a signature of such saturation is evident in Fig.~\ref{fig_3} in both devices.

Figure~\ref{fig_5} shows the variation of our fitted $I_{\text{h,s}}$ and $V_{\text{h,c}}$ with temperature for the devices A and B. It is seen that both parameters are essentially temperature independent, as expected by the FIT theory. We may evaluate the critical electric field, $E_{\text{h,c}}$, at the junction (nanocontact) from $V_{\text{h,c}}$ for the device A. Recall that the voltage mainly dropped at the two nanocontacts (because $R$\,$\simeq$\,2$R_{\text{c}}$). Taking a value $V_{\text{h,c}}$\,$\approx$\,43 mV (Fig.~\ref{fig_5}) and dividing it by $2w$\,$\approx$\,30 nm (Table~\ref{table_1}), we obtain $E_{\text{h,c}}$\,$\approx$\,1.4$\times$$10^6$ V/m. Theoretically, the FIT model predicts a critical electric field given by $E_{\text{h,c}}$\,=\,$4 \phi_0/ew$.\cite{ShengPRL78} (This corresponds to a critical voltage a few times of the junction barrier height.) Substituting the $\phi_0$ and $w$ values from Table~\ref{table_1}, we obtain $E_{\text{h,c}}$\,$\approx$\,5$\times$$10^5$ V/m. This agreement again supports the self-consistency of the FIT model. We mention in passing that the microscopic processes of the FIT conduction mechanism have recently been theoretically reexamined by considering the electronic wave transmission through long, narrow channels.\cite{XiePRB09}

\section{Conclusion}

In two-probe individual metallic NW devices and under appropriate conditions, the measured device resistance is dominated by the two nanocontact resistances. These nanocontact resistances are found to be well described by
the fluctuation-induced tunneling conduction theory. The parameters characterizing the nanocontacts can be quantitatively extracted. Taken together with our previous and complementary work (Ref.~\onlinecite{LinNT08}), our study demonstrates that the physical properties of electronic nanocontacts can be reliably inferred from the FIT model. This information is useful for the characterizations and implementation of nanoelectronic devices.

\begin{acknowledgements}

The authors are grateful to F. R. Chen and J. J. Kai for providing us with the RuO$_2$ NWs used in this study, and for P. Sheng and H. Xie for valuable discussions. This work was supported by the Taiwan National Science Council through Grant No. NSC 99-2120-M-009-001 and by the MOE ATU Program.

\end{acknowledgements}


\begin{thebibliography}{99}

\bibitem{1D} Z. M. Wang, \textit{One-Dimensional Nanostructures} (Springer, New York, 2010).

\bibitem{WangAPL04} H. Wang, J. Wang, M. Tian, L. Bell, E. Hutchinson, M. M. Rosario, Y. Liu, A. Amma, and T. Mallouk, Appl. Phys. Lett. {\bf 84}, 5171 (2004).

\bibitem{MarziJAP04} G. De Marzi, D. Iacopino, A. J. Quinn, and G. Redmond, J. Appl. Phys. {\bf 96}, 3458 (2004).

\bibitem{LinNT08} Y. H. Lin, S. P. Chiu, and J. J. Lin, Nanotechnology {\bf 19}, 365201 (2008).

\bibitem{ChiuNT09} S. P. Chiu, H. F. Chung, Y. H. Lin, J. J. Kai, F. R. Chen, and J. J. Lin, Nanotechnology {\bf 20}, 105203 (2009).

\bibitem{ShengPRL78} P. Sheng, E. K. Sichel, and J. I. Gittleman, Phys. Rev. Lett. {\bf 40}, 1197 (1978).

\bibitem{ShengPRB80} P. Sheng, Phys. Rev. B {\bf 21}, 2180 (1980).

\bibitem{Muster00} J. Muster, G. T. Kim, V. Krsti\'{c}, J. G. Park, Y. W. Park, S. Roth, and M. Burghard, Adv. Mater. {\bf 12}, 420 (2000).

\bibitem{ToimilMolares03} M. E. Toimil Molares, E. M. H\"{o}hberger, Ch. Schaeflein, R. H. Blick, R. Neumann, and C. Trautmann, Appl. Phys. Lett. {\bf 82}, 2139 (2003).

\bibitem{Gu01} G. Gu, M. Burghard, G. T. Kim, G. S. D\"{u}sberg, P. W. Chiu, V. Krstic, S. Roth, and W. Q. Han, J. Appl. Phys. {\bf 90}, 5747 (2001).

\bibitem{Nagashima10} K. Nagashima, T. Yanagida, A. Klamchuen, M. Kanai, K. Oka, S. Seki, and T. Kawai, Appl. Phys. Lett. {\bf 96}, 073110 (2010).

\bibitem{footnote1} A thin amorphous RuO$_2$ layer may exist on the surface of our NW. Nevertheless, amorphous RuO$_2$ retains a metallic behavior similar to that of its single-crystal form (see Ref.~\onlinecite{WangAPL04}). Its resistance, therefore, is immaterial as well.

\bibitem{footnote2} For example, an adsorbate layer introduced via the solution containing the NWs or an electron-beam contamination layer due to repetitive SEM inspections may give rise to this tunnel barrier.

\bibitem{footnote3} Unfortunately, due to an incidental burnout occurred during one of the $I$-$V$ curve sweeps, we did not obtain data points below 100 K for the device B.

\bibitem{footnote4} Note that in the FIT model the two metallic regions forming the tunnel junction are still large enough such that the Coulomb blockade effect is of no concern.

\bibitem{Wang94} Z. H. Wang, M. S. Dresselhaus, G. Dresselhaus, K. A. Wang, and P. C. Eklund, Phys. Rev. B {\bf 49}, 15890 (1994).

\bibitem{Fisher04} B. Fisher, K. B. Chashka, L. Patlagan, and G. M. Reisner, Phys. Rev. B {\bf 70}, 205109 (2004).

\bibitem{Paschen95} S. Paschen, M. N. Bussac, L. Zuppiroli, E. Minder, and B. Hilti, J. Appl. Phys. {\bf 78}, 3230 (1995).

\bibitem{Mandal11} G. Mandal, V. Srinivas, and V. V. Rao, J. Nanosci. Nanotechnol. {\bf 11}, 2570 (2011).

\bibitem{AhnJMC10} S. Ahn, Y. Kim, S. Beak, S. Ishimoto, H. Enozawa, E. Isomura, M. Hasegawa, M. Iyoda, and Y. Park, J. Mater. Chem. {\bf 20}, 10817 (2010).

\bibitem{KimSM01} G. T. Kim, S. H. Jhang, J. G. Park, Y. W. Park, and S. Roth, Synth. Met. {\bf 117}, 123 (2001).

\bibitem{SimmonsJAP63} J. G. Simmons, J. Appl. Phys. {\bf 34}, 1793 (1963).

\bibitem{XiePRB09} H. Xie and P. Sheng, Phys. Rev. B {\bf 79}, 165419 (2009).

\end{thebibliography}
\end{document}